\documentclass[draftcls,journal,onecolumn]{IEEEtran}

\usepackage{amsmath}
\usepackage{amsthm}
\usepackage{amssymb}
\usepackage{graphicx}
\usepackage[nolists]{endfloat}

\newtheoremstyle{default}% name
  {3pt}%      Space above
  {3pt}%      Space below
  {\itshape}%         Body font
  {}%         Indent amount (empty = no indent, \parindent = para indent)
  {\MakeUppercase{}}% Thm head font
  {.}%        Punctuation after thm head
  {.5em}%     Space after thm head: " " = normal interword space;
  {}%         Thm head spec (can be left empty, meaning `normal')

\theoremstyle{plain}
\newtheorem{theorem}{Theorem}

\graphicspath{{../../Pictures/}}

\title{Low-Signal-Energy Asymptotics of Capacity and Mutual Information for the Discrete-Time Poisson Channel}
\author{Alfonso Martinez%
  \thanks{A.\ Martinez is with Centrum Wiskunde \& Informatica,
  The Netherlands. % Tel: +31-40-2475742, Fax: +31-40-2466508.
  e-mail: alfonso.martinez@ieee.org.}}

\newcommand{\units}[2]{#1\textrm{\thinspace #2}}

\newcommand{\zz}{y}
\newcommand{\zzStar}{y^*}

\newcommand{\xx}{s}

\newcommand{\sss}{x}

\newcommand{\ww}{z}

\newcommand{\pmf}{P}

\newcommand{\chan}{Q}
\newcommand{\ew}{\varepsilon_{n}}
\newcommand{\es}{\varepsilon_{s}}

\newcommand{\eb}{\varepsilon_b}

\newcommand{\ebMin}{\varepsilon_{b,\text{min}}}
\newcommand{\ebZ}{\varepsilon_{b,0}}

\DeclareMathOperator{\snr}{SNR}

\newcommand{\capPois}{\text{C}}
\newcommand{\capERC}{\text{C}}

\newcommand{\pzx}{Q(y|x)}
\newcommand{\pzxp}{Q(y|x')}

\newcommand{\cUnitCost}{\text{C}_1}

\newcommand{\Xcal}{\mathcal{X}}

\newcommand{\muOne}{\mu_1}
\newcommand{\muTwo}{\mu_2}

\DeclareMathOperator{\Ord}{O}
\DeclareMathOperator{\ord}{o}

\newcommand{\ca}{c_1}
\newcommand{\cb}{c_2}

\graphicspath{{../../Pictures/}}

\begin{document}
\maketitle

\begin{abstract}
The first terms of the low-signal-energy asymptotics for the mutual information in the discrete-time Poisson channel are derived and compared to an asymptotic expression of the capacity. In the presence of non-zero additive noise (either Poisson or geometric), the mutual information is concave at zero signal-energy and the minimum energy per bit is not attained at zero capacity. Fixed signal constellations which scale with the signal energy do not attain the minimum energy per bit. The minimum energy per bit is zero when additive Poisson noise is present and $\ew\log 2$ when additive geometric noise of mean $\ew$ is present.
\end{abstract}

\section{Motivation and Notation}

In the complex-valued Gaussian channel with signal-to-noise ratio $\snr$ the mutual information of very general constellations (e.\ g.\ zero-mean with uncorrelated real and imaginary parts each of energy~$\frac{1}{2}$ \cite{prelov04:secondOrderAsymptoticsMutualInformation}) has the same low-$\snr$ asymptotics as the channel capacity, namely $\snr - \frac{1}{2}\snr^2 + \ord(\snr^2)$. These constellations also attain the minimum bit-energy-to-noise-variance ratio of \units{-1.59}{dB} at vanishing $\snr$.
A natural question concerns the extent to which this universality extends to other common channel models. We consider here the discrete-time Poisson channel, frequently used to represent optical communication channels, and quantity the gap between the channel capacity and the mutual information for fixed signal constellations. As a by-product of our analysis, we also determine the asymptotic form of the capacity at vanishing signal energy.

Consider a memoryless channel with input $X$ and output $Y$ given by the sum
\begin{equation}
  Y = S(X) + Z
\end{equation}
of a noise $Z$ and a signal component $S(X)$, itself a function of the input $X$. The input $X$ is a non-negative real number (i.\ e.\ it has units of energy), drawn from a unit-energy set $\Xcal$ according to a probability distribution $\pmf(x)$.
We let $S(X)$ be distributed according to a Poisson distribution of parameter $\es X$, where $\es$ is an average signal energy.
The output components $S(X)$, $Z$, and $Y$ are nonnegative integers. 

We study three channel models: noiseless, with $Z = 0$; additive Poisson noise, where $Z$ follows a Poisson distribution of mean $\ew > 0$;
and additive geometric noise, with $Z$ distributed according to a geometric distribution of mean $\ew > 0$. %; and additive Poisson noise, where $Z$ follows a geometric distribution of mean $\ew > 0$. 
With additive Poisson noise the channel transition probability, denoted by $\pzx$, is given by
\begin{align}
\pzx&=e^{-(\es\sss+\ew)}\frac{(\es\sss+\ew)^{\zz}}{\zz!} , \label{eq:chanzs-DTP}
\end{align}
where $\ew \geq 0$. For the channel with geometric noise, we have
\begin{align}
\pzx&=\sum_{l=0}^{\zz}\frac{e^{-\sss}}{1+\ew}\biggl(\frac{\ew}{1+\ew}\biggr)^{\zz} \frac{\Bigl(\sss\bigl(1+\tfrac{1}{\ew}\bigr)\Bigr)^{l}}{l!}. \label{eq:chanzs-AEQ}
\end{align}
Remark that the model with additive geometric noise arises in representations of electromagnetic radiation as a photon gas \cite{martinez08:informationRatesRadiationPhotonGas}.

In this letter, we compute the minimum energy per bit for these models. We also study the asymptotics of the mutual information $I\bigl(X;S(X)+Z\bigr)$ at low $\es$ and compare them with the channel capacity $\capERC(\es)$ at energy $\es$. 
The main results are presented in the next section; the proofs can be found in the appendices.

%. With non-zero geometric noise the minimum bit-energy-to-noise ratio $\eb/\ew$ is \units{-1.59}{dB}, as in the Gaussian channel. With Poisson noise the minimum energy per bit is known to be zero \cite{}. 

%We also study the asymptotics of the mutual information $I\bigl(X;S(X)+Z\bigr)$ at low $\es$ and compare them with the channel capacity $\capERC(\es)$ at energy $\es$. General signal constellations do not attain the minimum energy-per-bit at vanishing signal energy. An exception is flash signalling, whose minimum energy-per-bit is close to the best known upper bound to the capacity of the DTP channel \cite{}. In this case, the behaviour is consistent with a capacity expansion of the form $\capERC(\es) = \es\log\tfrac{1}{\es} + O(\es)$.  Moreover, in the presence of non-zero additive noise (either Poisson or geometric), the mutual information is concave at zero signal-energy and the minimum energy per bit is not attained at zero capacity. The main results are presented in the next section; the proofs can be found in the appendices.

\section{Main Results and Discussion}

\subsection{Capacity Asymptotics and Capacity per Unit Cost}
A closed-form expression for the capacity $\capPois(\es)$ of the discrete-time Poisson channel is not known. For the noiseless channel (i.\ e.\ with $Z = 0)$, the best known firm upper bound was derived in \cite{martinez07:spectralEfficiencyOpticalDirectDetection}, and is given by
\begin{align}\label{eq:upBound-Poisson}
    \capPois(\es) &\leq \log\Biggl(\biggl(1+\frac{\sqrt{2e}-1}{\sqrt{1+2\es}}\biggr)\frac{\bigl(\es+\frac{1}{2}\bigr)^{\es+\frac{1}{2}}}{\sqrt{e}\es^{\es}}\Biggr). 
\end{align}  

As for lower bounds, binary modulation attains a high mutual information at low values of $\es$. Specifically, let the symbols be located at $x = 0$ and $x = 1/p$, and be respectively used with probabilities $1-p$ and $p$. We denote the mutual information attained by such modulation by $I_{\rm b}(p)$. A trite computation gives
\begin{align}\label{eq:mInfFlash}
    I_{\rm b}(p) &= -(p-pe^{-\frac{\es}{p}})\log p - \es e^{-\frac{\es}{p}} -(1-p+pe^{-\frac{\es}{p}})\log(1-p+pe^{-\frac{\es}{p}}).
%    I_{\rm b}(p) &= - \es e^{-\frac{\es}{p}} -\log\left(p^{p-pe^{-\frac{\es}{p}}}(1-p+pe^{-\frac{\es}{p}})^{1-p+pe^{-\frac{\es}{p}}}\right).
\end{align}  

Setting $p = \es$, and using the Taylor expansion of the logarithm $\log(1+x)$ around $x = 0$ in both Eq.~\eqref{eq:upBound-Poisson} and Eq.~\eqref{eq:mInfFlash}, we have that
\begin{align}\label{eq:sandwichCapacity}
    -(1-e^{-1})\es\log\es + \Ord(\es) \lesssim \capPois(\es) \lesssim -\es\log\es + \Ord(\es).
\end{align}  
%Similarly, setting $p = \es$ in $I_{\rm b}(p)$, we easily obtain
%\begin{align}
%    \capPois(\es) &\gtrsim -(1-e^{-1})\es\log\es + \Ord(\es).
%\end{align}  
Therefore, the capacity $\capPois(\es)$ of the noiseless discrete-time Poisson channel behaves as $\Ord(-\es\log\es)$ at vanishing $\es$. 

Moreover, a similar reasoning shows that flash signalling with $p = -\es\log\es$ (for $\es \ll 1$) asymptotically behaves as
\begin{equation}\label{eq:lowBoundAsymp}
  I_{\rm b}(-\es\log\es) = -\es\log\es + \ord(-\es\log\es)
\end{equation}
for low $\es$. Combining Eqs.~\eqref{eq:sandwichCapacity} and~\eqref{eq:lowBoundAsymp}, we obtain the following
\begin{theorem}\label{thm:capPoisLowEs}
For vanishing $\es$ the capacity $\capPois(\es)$ behaves as 
\begin{equation}
  \capPois(\es) = -\es\log\es + \ord(-\es\log\es).
\end{equation}
\end{theorem}
This result complements the asymptotic behaviour for very large values of $\es$, which was established in \cite{lapidoth03:boundsCapacityPoisson,martinez07:spectralEfficiencyOpticalDirectDetection},
\begin{equation}
  \capPois(\es) = \frac{1}{2}\log\es +\ord(\log\es).
\end{equation}

Observe that Eq.~\eqref{eq:sandwichCapacity} implies that the capacity per unit energy $\cUnitCost$, 
\begin{equation}\label{eq:deft-cUnitCost}
    \cUnitCost = \sup_{\es} \frac{\capERC(\es)}{\es}.
\end{equation}
is infinite for the discrete-time Poisson channel. This well-known result had been obtained by Verd{\'u} \cite{verdu90:capacityUnitCost} by exploiting a simple formula for $\cUnitCost$ in channels which have a zero-energy symbol ($x = 0$ in our case), namely
\begin{align}
    \cUnitCost &= \sup_{\sss} \frac{D\bigl(\chan(\zz|\sss)||\chan(\zz|\sss=0)\bigr)}{\es \sss},
\end{align}
where $D\bigl(\chan(\zz|\sss)||\chan(\zz|0)\bigr)$ is the divergence between the transition probabilities $\chan(\zz|\sss)$ for arbitrary input $\sss$ and zero input $\sss = 0$. 

By definition, the minimum energy per bit $\ebMin = \inf_{\es} \frac{\es}{\capERC(\es)}$, where the capacity is measured in bits, is given by
%If the capacity is in nats, it is clear from the definitions that
%\begin{equation}\label{eq:deft-ebMin}
    $\ebMin = \frac{\log 2}{\cUnitCost}$.
%\end{equation}

Applied to the family of channels we consider, we  have
\begin{theorem}
In the absence of additive noise, i.\ e.\, for $Z = 0$, or in the presence of additive Poisson noise, the minimum energy per bit is $\ebMin = 0$. Equivalently the capacity per unit cost is $\cUnitCost = \infty$. 

With additive geometric noise, the minimum energy per bit is $\ebMin = \ew\log 2$ and the capacity per unit cost is $\cUnitCost = \units{\ew^{-1}}{nats}$. 
\end{theorem}
The latter result is new. Remarkably, the minimum energy per bit has the same form as in the Gaussian channel, for a minimum ratio $\eb/\ew$ of \units{-1.59}{dB}.
\begin{proof}
The proof can be found in Appendix~\ref{app:cUnitCost}.
\end{proof}

\subsection{Mutual Information Asymptotics}
We now move on to study the asymptotics of the mutual information $I\bigl(X;S(X)+Z\bigr)$ (in nats) at low $\es$ for fixed unit-energy constellations $\Xcal$. The output $S(X)$ is distributed according to a Poisson distribution of parameter $\es X$.
We determine the first two coefficients $\ca$ and $\cb$ in the Taylor expansion around $\es = 0$, that is
\begin{equation}\label{eq:expansionMInf}
	I\bigl(X;S(X)+Z\bigr) = \ca \es + \cb \es^2 + \ord(\es^2).
\end{equation}
Also, we define the energy per bit $\eb$ as
%\begin{equation}
	$\eb = \frac{\es}{I(X;Y)}\log 2$.
%\end{equation}

Denoting the first- and second-order moments of the constellation by $\muOne$ (we often have $\muOne = 1$) and $\muTwo$ respectively, we have
\begin{theorem}
In the absence of additive noise, i.\ e.\, for $Z = 0$, the mutual information behaves at low $\es$ as Eq.~\eqref{eq:expansionMInf} with
\begin{align}\label{eq:cmAsymp-DTP}
    \ca = \sum_{\sss\in\Xcal}\pmf(\sss)\sss\log \frac{\sss}{\muOne}, \quad \cb = \frac{1}{2}\biggl(\muTwo-\muOne^2- \muTwo\log \frac{\muTwo}{\muOne^2}\biggr) .
\end{align}

As long as $\ew > 0$, regardless of whether $Z$ has a Poisson or a geometric distribution, the coefficients in Eq.~\eqref{eq:expansionMInf} are
\begin{align}\label{eq:cmAsymp-AEQ}
    \ca = 0, \quad \cb = \frac{1}{2}\frac{(\muTwo-\muOne^2)}{\ew}.
\end{align}
\end{theorem}
\begin{proof}
Eq.~\eqref{eq:cmAsymp-DTP} is proved in Appendix~\ref{app:cmExpansionDTP}, Eq.~\eqref{eq:cmAsymp-AEQ}  in Appendices~\ref{app:cmExpansionDTPb} and~\ref{app:cmExpansionAEQ} for Poisson and geometric noise respectively.
\end{proof}

\subsection{Discussion}

\begin{figure}[htb]
  \centering
%  \strictpagechecktrue
%  \begin{adjustwidth*}{-0.5cm}{-1.3cm}
%    \begin{minipage}[b]{0.575\textwidth}
%        \subfloat[Low $\es$.]{\label{fig:t20070112-esa}\includegraphics{t20070112-es}}
%    \end{minipage}
%    \begin{minipage}[b]{0.575\textwidth}
%        \subfloat[Moderate-to-high $\es$.]{\label{fig:t20070112-ebLog}\includegraphics{t20070112-ebLog}}
  \includegraphics[width=\columnwidth]{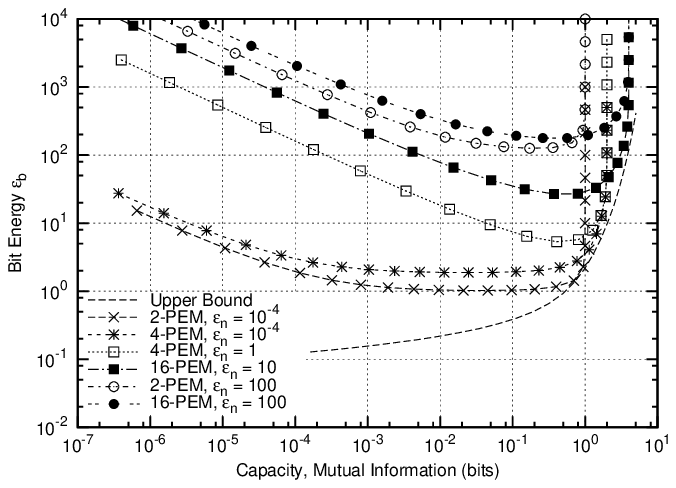}
%    \end{minipage}
%  \end{adjustwidth*}
  \caption{Bit energy $\eb$ as a function of the mutual information for uniform $2^m$-PEM and varying $\ew$ (geometric noise).}
  \label{fig:t20070112-ebLog}
\end{figure}

In the presence of non-zero additive noise, the mutual information is concave at zero signal-energy (because $\ca = 0$, $\cb > 0$) and the minimum energy per bit is not attained at zero capacity. This effect can be seen in Fig.~\ref{fig:t20070112-ebLog}, which depicts the energy per bit $\eb$ as a function of the mutual information for several positive values of $\ew$. The modulation depicted is uniform pulse-energy modulation (PEM), e.\ g.\ $2^m$ points uniformly located between 0 and 2 with spacing $1/2^m$. General signal constellations do not attain the minimum energy-per-bit at vanishing signal energy. The determination of the minimum energy per bit attained by these modulations is an open problem.

\begin{figure}[htb]
  \centering
%  \strictpagechecktrue
%  \begin{adjustwidth*}{-0.5cm}{-1.3cm}
%    \begin{minipage}[b]{\columnwidth}
%        \subfloat[Capacity as a function of $\es$.]{\label{fig:t20070215-snr}\includegraphics{t20070215-snr}}
%    \end{minipage}
%    \begin{minipage}[b]{\columnwidth}
%        \subfloat[$\eb$ as a function of the capacity.]{\label{fig:t20070215-bnr}\includegraphics{t20070215-bnr}}
        \includegraphics[width=\columnwidth]{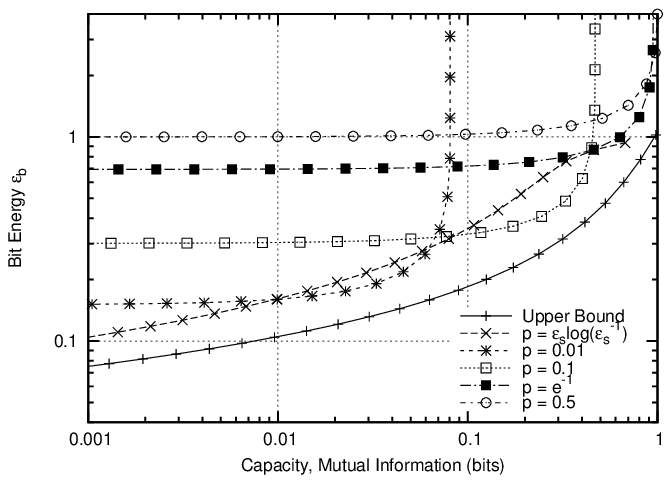}
%    \end{minipage}
%  \end{adjustwidth*}
  \caption{Bit energy $\eb$ as a function of the mutual information or the capacity (noiseless channel, $Z = 0$).}
  \label{fig:t20070215-bnr}
\end{figure}

Moreover, since there is no coefficient in $-\es\log\es$ in Eq.~\eqref{eq:expansionMInf}, these modulations do not attain the minimum energy per bit $\ebMin$. This is true even for the noiseless channel, for which $\ca \neq 0$. In this case, binary modulation at points $x=0$ and $x=1/p$ respectively used with probabilities $1-p$ and $p$, with fixed $p$, has coefficients $\ca$ and $\cb$
\begin{align}
    \ca =  -\log(p),\quad
    \cb =  \frac{1-p+\log(p)}{2p}.
\end{align}
In the limit $p\to 0$, $\ca\to\infty$ and the bit energy at zero capacity, $\ebZ$, approaches 0, the minimum energy per bit.
Fig.~\ref{fig:t20070215-bnr} depicts $\eb$ as a function of the mutual information for various fixed values of $p$ and for $p = -\es\log\es$. For comparison, we also include the value of $\eb$ corresponding to the upper bound in Eq.~\eqref{eq:upBound-Poisson}.
%The envelope of the family curves for $I(X;Y)$ with varying $p$ is close to the upper bound. Observe that this upper bound behaves $\capPois(\es)\simeq \es\log\es^{-1}$ for vanishing $\es$, and a similar decay seems to hold for the envelope of the curves with varying $p$. 
%Fig.~\ref{fig:t20070215-bnr} shows $\eb$ as a function of the capacity. 
Even though $\ebZ$ indeed approaches zero, it does so rather slowly. Also, the gap between the energy per bit corresponding to $I_{\rm b}(-\es\log\es)$ and $\ebMin$ closes slowly. Numerical evaluation shows that it is only for values of $\es$ below $10^{-307}$ (!) that $I_{\rm b}$ exceeds $0.99\cdot\es\log\es^{-1}$. Even though one eventually has $\capPois(\es) \simeq -\es\log\es$, convergence to the limit is very slow. This fact, together with the concave nature of the mutual information $I(X;Y)$ at zero $\es$ for nonzero additive noise, suggests that the asymptotic analysis of the capacity and the mutual information in the discrete-time Poisson channel fails to capture the key features of these quantities. This behaviour stands in contrast with the Gaussian channel, where asymptotic expansions give an accurate representation of the capacity and the mutual information \cite{verdu02:efficiencyWB}.

\appendices

\section{Capacity per Unit Energy}
\label{app:cUnitCost}

We first consider the case with additive Poisson noise.
Using Eq.~\eqref{eq:chanzs-DTP} for $\chan(\cdot|\cdot)$ and the definition of divergence, we have
\begin{align}
    D\bigl(\chan(\zz|\sss)||\chan(\zz|0)\bigr) %&= \sum_\zz \pzx\log\frac{e^{-\es \sss}(\es \sss+\ew)^\zz}{\ew^\zz} \\
%    &= \sum_\zz \pzx\left(-\es \sss + \zz\log\frac{\es \sss+\ew}{\ew}\right)
    &= -\es x + (\es x+\ew)\log\frac{\es \sss+\ew}{\ew}.
\end{align}
Hence,
\begin{align}
    \cUnitCost &= \sup_{\sss} \left(-1 + \left(1+\frac{\ew}{\es \sss}\right)\log\frac{\es \sss +\ew}{\ew}\right) = \infty.
\end{align}

We now consider the channel with additive geometric noise.
Using Eq.~\eqref{eq:chanzs-AEQ} for $\chan(\cdot|\cdot)$ and the definition of divergence, we have
\begin{align}
    D\bigl(\chan(\zz|\sss)||\chan(\zz|0)\bigr) %&= \sum_z \pzx\log\frac{\pzx}{Q(z|s=0)} \\
%    &= \sum_z \pzx\log\Biggl(e^{-\es s}\sum_{l=0}^z \Biggl(\es s\biggl(1+\frac{1}{\ew}\biggr)\Biggr)^{l}\frac{1}{l!}\Biggr) \\
    &= \sum_\zz \pzx\log\Biggl(e^{\frac{\es \sss}{\ew}}\sum_{l=0}^\zz e^{-\alpha}\frac{\alpha^{l}}{l!}\Biggr),
\end{align}
where $\alpha = \es \sss\bigl(1+\frac{1}{\ew}\bigr)$. Let us define $\pmf(l) = e^{-\alpha}\frac{\alpha^{l}}{l!}$ and the quantity $q(\zz) = \sum_{l=0}^\zz \pmf(l)$,
%\begin{equation}
%  q(\zz) = \sum_{l=0}^\zz e^{-\alpha}\frac{\alpha^{l}}{l!},
%\end{equation}
i.\ e.\ the cumulative distribution function of a Poisson random variable with mean $\alpha$.

Moving the exponential out of the logarithm, we obtain
\begin{align}
    D\bigl(\chan(\zz|\sss)||\chan(\zz|0)\bigr) &= \frac{\es \sss}{\ew}+\sum_\zz \pzx\log\bigl(q(\zz)\bigr).
\end{align}
Hence, the capacity per unit energy is given by
\begin{align}
    \cUnitCost &= \frac{1}{\ew} + \sup_{\sss} \frac{\sum_\zz \pzx\log\bigl(q(\zz)\bigr)}{\es \sss}. %    \\
%    &= \frac{1}{\ew} + \sup_{s} \frac{\sum_z e^{\frac{\es s}{\ew}}\frac{1}{1+\ew}\bigl(\frac{\ew}{1+\ew}\bigr)^{z}\sum_{l=0}^z \frac{e^{-\alpha}\alpha^{l}}{l!}\log\Bigl(\sum_{l=0}^z e^{-\alpha}\frac{\alpha^{l}}{l!}\Bigr)}{\es s} .
\end{align}
Since $q(\zz) \leq 1$, its logarithm is always non-positive, and
\begin{align}\label{eq:cUnitCost-AEQ-ub}
    \cUnitCost &\leq \frac{1}{\ew}.
\end{align}
The proof is completed by proving that
\begin{align}\label{eq:5.96}
    \lim_{\sss\to\infty} \frac{e^{\tfrac{\es\sss}{\ew}}}{\es \sss(1+\ew)}\Biggl(\sum_\zz \biggl(\frac{\ew}{1+\ew}\biggr)^{\zz} q(\zz)\log\bigl(q(\zz)\bigr)\Biggr) = 0,
\end{align}
where we expressed $\pzx$ as a function of $q(\zz)$. If this condition holds true, then Eq.~\eqref{eq:cUnitCost-AEQ-ub} becomes an equality.

In Eq.~\eqref{eq:5.96} we split the summation over $\zz$ into two parts, from 0 to $\zzStar = \lfloor\alpha\rfloor$, and from $\zzStar+1$ to infinity. In the first part, $e^{-\alpha}\frac{\alpha^\zz}{\zz!}$ is an increasing function in $\zz$, and therefore
\begin{equation}
  q(\zz) = \sum_{l=0}^\zz \pmf(l) \geq \sum_{l=0}^\zz \pmf(0) = (\zz+1)\pmf(0) = (\zz+1)e^{-\alpha}.
\end{equation}
Hence, the summation for $\zz \leq \zzStar$ is bounded as
\begin{align}
  \sum_{\zz=0}^{\zzStar} \biggl(\frac{\ew}{1+\ew}\biggr)^{\zz} q(\zz)\log\bigl(q(\zz)\bigr) \geq \sum_{\zz=0}^{\zzStar} \biggl(\frac{\ew}{1+\ew}\biggr)^{\zz} (\zz+1)e^{-\alpha}\bigl(\log(\zz+1)-\alpha\bigr).
\end{align}
And, multiplying by the exponential factor $e^{\tfrac{\es\sss}{\ew}}$, we have
\begin{align}
  \sum_{\zz=0}^{\zzStar} e^{-\es \sss}\biggl(\frac{\ew}{1+\ew}\biggr)^{\zz} (\zz+1)\log(\zz+1)-\sum_{\zz=0}^{\zzStar} e^{-\es \sss}\biggl(\frac{\ew}{1+\ew}\biggr)^{\zz} (\zz+1)\alpha.
\end{align}
Both summands vanish as $\zz\to\infty$. The second has the form
\begin{align}
  e^{-\es \sss}\sum_{\zz=0}^{\zzStar} \biggl(\frac{\ew}{1+\ew}\biggr)^{\zz} (\zz+1)\alpha,
\end{align}
which decays exponentially in $\sss$, since the sum satisfies
\begin{align}
  \sum_{\zz=0}^{\zzStar} \biggl(\frac{\ew}{1+\ew}\biggr)^{\zz} (\zz+1)\leq \sum_{\zz=0}^{\infty} \biggl(\frac{\ew}{1+\ew}\biggr)^{\zz} (\zz+1) = (1+\ew)^2,
\end{align}
and $e^{-\es \sss}\alpha(1+\ew)^2$ vanishes for large $x$. Similarly, the summation
\begin{align}
  \sum_{\zz=0}^{\zzStar} \biggl(\frac{\ew}{1+\ew}\biggr)^{\zz} (\zz+1)\log(\zz+1)
\end{align}
remains bounded, since it is the partial sum of a convergent series, with $n$-th coefficient $\beta^n(n+1)\log(n+1)$ and $\beta = \ew/(1+\ew) < 1$. This is verified by the checking the ratio test, as
\begin{align}
  \lim_{n\to\infty} \frac{\beta^n(n+1)\log(n+1)}{\beta^{n-1}n\log n} = \beta < 1.
\end{align}
Boundedness of the partial sum implies that, after multiplying times an exponential factor $e^{-\es \sss}$, the first summand vanishes as $\sss\to\infty$.

Next, we consider the remainder of the summation in Eq.~\eqref{eq:5.96},
\begin{align}
    \sum_{\zz=\zzStar+1}^\infty \biggl(\frac{\ew}{1+\ew}\biggr)^{\zz} q(\zz)\log\bigl(q(\zz)\bigr).
\end{align}
Clearly, $q(0) = e^{-\alpha} \leq q(\zz) \leq 1$ and therefore $-\alpha \leq \log q(\zz) \leq 0$, so each summand is negative and bounded by
\begin{align}
    \biggl(\frac{\ew}{1+\ew}\biggr)^{\zz} q(\zz)\log\bigl(q(\zz)\bigr) &\geq -\alpha\biggl(\frac{\ew}{1+\ew}\biggr)^{\zz} q(\zz) \geq -\alpha\biggl(\frac{\ew}{1+\ew}\biggr)^{\zz}.
\end{align}
Summing over $\zz$,
\begin{align}
    \sum_{\zz=\zzStar+1}^\infty \biggl(\frac{\ew}{1+\ew}\biggr)^{\zz} q(\zz)\log\bigl(q(\zz)\bigr) \geq -\alpha(1+\ew)\biggl(\frac{\ew}{1+\ew}\biggr)^{\zzStar+1}.
\end{align}

%Moreover, since $\bigl(\frac{\ew}{1+\ew}\bigr)^\zz$ is a decreasing function with $\zz$, this quantity is clearly upper bounded as
%\begin{align}
%    \sum_{\zz=\zzStar+1}^\infty \biggl(\frac{\ew}{1+\ew}\biggr)^{\zz} q(\zz)\log\bigl(q(\zz)\bigr) &\geq
%    \biggl(\frac{\ew}{1+\ew}\biggr)^{\zzStar+1} \sum_{\zz=\zzStar+1}^\infty q(\zz)\log\bigl(q(\zz)\bigr)\\
%    &\geq \biggl(\frac{\ew}{1+\ew}\biggr)^{\zzStar+1}\sum_{\zz=0}^\infty  q(\zz)\log\bigl(q(\zz)\bigr),
%\end{align}
%where we have exploited that extending the summation back to $\zz = 0$ only adds (negative) terms. The quantity $\sum_{\zz=0}^\infty  q(\zz)\log\bigl(q(\zz)\bigr)$ does not grow exponentially in $\sss$; we have verified numerically that is rather grows as $\Ord(\sqrt{\alpha})$ for large $\alpha$.
%
Using that  $\alpha = \es \sss\bigl(1+\frac{1}{\ew}\bigr)$ and taking into account the denominator $\es \sss(1+\ew)$ in Eq.~\eqref{eq:5.96}, we must study the behaviour of
\begin{align}
  -\frac{\ew+1}{\ew}e^{\tfrac{\es\sss}{\ew}}\biggl(\frac{\ew}{1+\ew}\biggr)^{\zzStar+1} = -\frac{\ew+1}{\ew}e^{\tfrac{\es\sss}{\ew}-(\zzStar+1)\log\bigl(1+\frac{1}{\ew}\bigr)}\label{eq:5.106}
\end{align}
as $\sss\to\infty$. By construction, $\zzStar + 1 > \alpha$, and therefore
\begin{align}\label{eq:5.155}
  \frac{\es\sss}{\ew}-(\zzStar+1)\log\biggl(1+\frac{1}{\ew}\biggr) &<
  \frac{\es\sss}{\ew}-\es \sss\biggl(1+\frac{1}{\ew}\biggr)\log\biggl(1+\frac{1}{\ew}\biggr) \\ &= \es\sss\Biggl(\frac{1}{\ew}-\biggl(1+\frac{1}{\ew}\biggr)\log\biggl(1+\frac{1}{\ew}\biggr)\Biggr).
%  &\leq 0
\end{align}
Since $t \leq (1+t)\log(1+t)$ for $t> 0$, a fact which follows from the inequality $\log(1+t)\leq t$, the left-hand side of Eq.~\eqref{eq:5.155} is strictly upper bounded by a function $ax$, with $a < 0$. Hence, the function in Eq.~\eqref{eq:5.106} vanishes exponentially as $\sss \to\infty$, and so does the term
\begin{align}
    \frac{e^{\tfrac{\es\sss}{\ew}}}{\es \sss(1+\ew)}\sum_{\zz=\zzStar+1}^\infty \biggl(\frac{\ew}{1+\ew}\biggr)^{\zz} q(\zz)\log\bigl(q(\zz)\bigr).
\end{align}
This proves the limit in Eq.~\eqref{eq:5.96} and that Eq.~\eqref{eq:cUnitCost-AEQ-ub} holds with equality.

\section{Asymptotics at Low $\es$ for $Z = 0$}
\label{app:cmExpansionDTP}

The mutual information is given by
\begin{align}
    I(X;Y) &= - \sum_\sss\pmf(\sss) \sum_{\zz=0}^\infty \pzx \log \Biggl(\sum_{\sss'\in\Xcal}\pmf(\sss'){e^{\es (\sss-\sss')}\biggl(\frac{\sss'}{\sss}\biggr)^\zz}\Biggr). \label{eq:314}
\end{align}

Using the Taylor expansion of the exponential
    $e^{t} = 1 + t + \frac{1}{2}t^2 + \Ord(t^3)$,
we notice that there are only three possible channel outputs to order $\es^3$, namely
\begin{align}
    &\zz = 0,\quad \pzx = 1-\es \sss +\tfrac{1}{2}\es^2 \sss^2 + \Ord(\es^3) \label{eq:326}\\
    &\zz = 1,\quad \pzx = \es \sss -\es^2\sss^2 +\Ord(\es^3) \label{eq:327}\\
    &\zz = 2,\quad \pzx = \tfrac{1}{2}\es^2 \sss^2+\Ord(\es^3)\label{eq:328}\\
    &\zz > 2,\quad \pzx = \Ord(\es^3).
\end{align}
Since each of these cases behaves differently, we examine them separately.

We rewrite the variable in the $\log(\cdot)$ in Eq.~\eqref{eq:314} with the appropriate approximation. When the output is $\zz = 0$, the variable is
\begin{align}
    \sum_{\sss'\in\Xcal}\pmf(\sss')e^{\es (\sss-\sss')} &= \sum_{\sss'\in\Xcal}\pmf(\sss')\Bigl(1+\es (\sss-\sss') +\tfrac{1}{2}\es^2 (\sss-\sss')^2 + \Ord(\es^3)\Bigl) \\
    &= 1 + \es(\sss-\muOne) + \tfrac{1}{2}\es^2 (\sss^2+\muTwo-2\sss\muOne) + \Ord(\es^3).
\end{align}
Taking logarithms, and using the formula
%\begin{equation}
    $\log(1+t) = t -\frac{1}{2}t^2+ \Ord(t^3)$,
%\end{equation}
we obtain
\begin{align}
    \es(\sss-\muOne) &+ \tfrac{1}{2}\es^2 (\sss^2+\muTwo-2\sss\muOne) -\tfrac{1}{2}\es^2(\sss^2+\muOne^2-2\sss\muOne)+ \Ord(\es^3) \\
    &= \es(\sss-\muOne) + \tfrac{1}{2}\es^2 \bigl(\muTwo-\muOne^2\bigr)+ \Ord(\es^3)\label{eq:321}.
\end{align}

For $\zz = 1$, the variable in the logarithm in Eq.~\eqref{eq:314} is
\begin{align}
    \sum_{\sss'\in\Xcal}\pmf(\sss') e^{\es (\sss-\sss')}\frac{\sss'}{\sss} &= \frac{1}{\sss}\sum_{\sss'\in\Xcal}\pmf(\sss')\sss' \Bigl(1+\es (\sss-\sss')+ \Ord(\es^2)\Bigl)\\
%    &= \frac{1}{\sss} \Bigl(\muOne+\es \bigl(\sss\muOne-\muTwo\bigr)+ \Ord(\es^2)\Bigl)\\
    &= \frac{\muOne}{\sss} \Bigl(1+\frac{\es}{\muOne}\bigl(\sss\muOne-\muTwo\bigr)+ \Ord(\es^2)\Bigl).
\end{align}
Taking logarithms, and using the Taylor expansion, we get
\begin{align}
    \log \frac{\muOne}{\sss} +\frac{\es}{\muOne}\bigl(\sss\muOne-\muTwo\bigr)+ \Ord(\es^2).\label{eq:323}
\end{align}
We will later verify that no higher-order terms are required.

At last, for $\zz = 2$, the variable in the logarithm in Eq.~\eqref{eq:314} is
\begin{align}
    \sum_{\sss'\in\Xcal}\pmf(\sss') e^{\es (\sss-\sss')}\frac{\sss'^2}{\sss^2} &= \frac{1}{\sss^2}\sum_{\sss'\in\Xcal}\pmf(\sss')\sss'^2 \Bigl(1+ \Ord(\es)\Bigl)\\
%    &= \frac{1}{\sss^2} \Bigl(\muTwo+ \Ord(\es)\Bigl)\\
    &= \frac{\muTwo}{\sss^2} \bigl(1+ \Ord(\es)\bigl).
\end{align}
Again, taking logarithms, and using the Taylor expansion, we get
\begin{align}
    \log \frac{\muTwo}{\sss^2} + \Ord(\es).\label{eq:323-bis}
\end{align}
Later, we will verify that no higher-order terms are required.

After carrying out the averaging over $\zz$, we first combine Eqs.~\eqref{eq:321},~\eqref{eq:323} and~\eqref{eq:323-bis} with the probabilities in Eqs.~\eqref{eq:326}--\eqref{eq:328} and then group all terms up to $\Ord(\es^3)$ to derive
\begin{align}
    &\Bigl(1-\es \sss +\tfrac{1}{2}\es^2 \sss^2\Bigr)\Bigl(\es(\sss-\muOne) + \tfrac{1}{2}\es^2 \bigl(\muTwo-\muOne^2\bigr)\Bigr) + \notag\\
    &\quad +\bigl(\es \sss -\es^2\sss^2\bigr)\biggl(\log \frac{\muOne}{\sss} +\frac{\es}{\muOne}\bigl(\sss\muOne-\muTwo\bigr)\biggr) +\tfrac{1}{2}\es^2 \sss^2\log \frac{\muTwo}{\sss^2}+ \Ord(\es^3)\\
    &= \es(\sss-\muOne) + \tfrac{1}{2}\es^2 \bigl(\muTwo-\muOne^2\bigr) -\es^2(\sss^2-\sss\muOne) +\es \sss\log \frac{\muOne}{\sss} \notag \\
    &\quad -\es^2\sss^2\log \frac{\muOne}{\sss} + \es^2 \frac{1}{\muOne}\bigl(\sss^2\muOne-\sss\muTwo\bigr)+\tfrac{1}{2}\es^2 \sss^2\log \frac{\muTwo}{\sss^2}+ \Ord(\es^3).
%    &= -\log \pmf(\sss) - \es + \es \sss +\es \sss\log \pmf(\sss) - \es \sss\log \pmf(\sss) - \es \sss\log \sss + \Ord(\es^2) \\
%    &= -\log \pmf(\sss) - \es + \es \sss  - \es \sss\log \sss + \Ord(\es^2)
\end{align}
%
%\begin{align}
%    &= \frac{\sum_{\sss'\in\Xcal}{\pzxp}\pmf(\sss')}{\pzx\pmf(\sss)} \\
%    &= \frac{\sum_{\sss'\in\Xcal}\bigl({\es \sss'} + \Ord(\es^2)\bigr)\pmf(\sss')}{\bigl({\es s} + \Ord(\es^2)\bigr)\pmf(\sss)} \\
%    &= \frac{1 + \Ord(\es)}{\bigl(s + \Ord(\es)\bigr)\pmf(\sss)} \\
%    &= \frac{1}{s\pmf(\sss)} + \Ord(\es) \\
%    &= \frac{1}{s\pmf(\sss)}\bigl(1+ s\pmf(\sss)\Ord(\es)\bigr).
%\end{align}
%
%Taking logarithms, we have
%\begin{align}\label{eq:221}
%    -\log \pmf(\sss) &-\log s +\log \bigl(1+ s\pmf(\sss)\Ord(\es)\bigr) \\
%    -\log \pmf(\sss) &-\log s + s\pmf(\sss)\Ord(\es)\label{eq:224}.
%\end{align}
%
%The remaining steps are the expectation over $s$ and $\zz$.
%Starting with $\zz$, we combine both equations to obtain
%\begin{align}
%    (1-&\es s)\bigl(-\log \pmf(\sss) - \es + \es s + \Ord(\es^2)\bigr) \notag \\ &\qquad\qquad+ \es s\bigl(-\log \pmf(\sss) -\log s + s\pmf(\sss)\Ord(\es)\bigr) \\
%    &= -\log \pmf(\sss) - \es + \es s +\es s\log \pmf(\sss) - \es s\log \pmf(\sss) - \es s\log s + \Ord(\es^2) \\
%    &= -\log \pmf(\sss) - \es + \es s  - \es s\log s + \Ord(\es^2)
%\end{align}
%with the obvious simplifications. Note that the term $\Ord(\es)$ in Eqs.~\eqref{eq:221} and~\eqref{eq:224} does not lead to any difficulties.
The expectation over $\sss$ is straightforward, and gives the desired $I(X;Y)$.
%\begin{align}
%    \Hh(S) - \es \sum_{\sss\in\Xcal}\pmf(\sss)\sss\log \sss + \Ord(\es^2).
%\end{align}
%And, finally,
%\begin{align}
%    \capCM &= \es \sum_{\sss\in\Xcal}\pmf(\sss)\sss\log \frac{\sss}{\muOne} +\frac{1}{2}\es^2 \biggl(\muTwo-\muOne^2- \muTwo\log \frac{\muTwo}{\muOne^2}\biggr)  + \Ord(\es^3).
%\end{align}

\section{Asymptotics at Low $\es$ for Poisson Noise}
\label{app:cmExpansionDTPb}

The mutual information is given by
\begin{align}
    I(X;Y) &= - \sum_\sss\pmf(\sss) \sum_{\zz=0}^\infty \pzx \log \left(\sum_{\sss'\in\Xcal}\pmf(\sss'){e^{\es (\sss-\sss')}\frac{\Bigl(1+\frac{\es\sss'}{\ew}\Bigr)^\zz}{\Bigl(1+\frac{\es\sss}{\ew}\Bigr)^\zz}}\right). \label{eq:314b}
\end{align}

Using the Taylor expansion of the exponential,
%    $e^{t} = 1 + t + \frac{1}{2}t^2 + \Ord(t^3)$,
and neglecting terms of order higher than $\es^3$, the channel output law is given by
\begin{align}
  \pzx = \frac{\ew^\zz}{\zz!}e^{-\ew}\left(1+\es\sss\left(\frac{\zz}{\ew}-1\right)+\frac{(\es\sss)^2}{2}\left(\frac{\zz(\zz-1)}{\ew^2}+1-\frac{2\zz}{\ew}\right)\right)+ \Ord(\es^3).
\end{align}

We next examine the logarithm in Eq.~\eqref{eq:314b}. First, the Taylor expansions of $(1+t)^{y}$ and $(1+t)^{-y}$ yield
\begin{align}
    \frac{\Bigl(1+\frac{\es\sss'}{\ew}\Bigr)^\zz}{\Bigl(1+\frac{\es\sss}{\ew}\Bigr)^\zz} %&= \left(1+\zz\frac{\es\sss'}{\ew}+\frac{\zz(\zz-1)}{2}\frac{(\es\sss')^2}{\ew^2}+\Ord(\es^3)\right)\left(1-\zz\frac{\es\sss}{\ew}+\frac{\zz(\zz+1)}{2}\frac{(\es\sss)^2}{\ew^2}+\Ord(\es^3)\right) \\
%    &= 1+\zz\frac{\es\sss'}{\ew}+\frac{\zz(\zz-1)}{2}\frac{(\es\sss')^2}{\ew^2} -\zz\frac{\es\sss}{\ew}-\zz\frac{\es\sss}{\ew}\zz\frac{\es\sss'}{\ew} + \frac{\zz(\zz+1)}{2}\frac{(\es\sss)^2}{\ew^2} + \Ord(\es^3) \\
    &= 1+\frac{\es}{\ew}\zz(\sss'-\sss)+\frac{\es^2}{2\ew^2}\bigl(\zz^2(\sss'-\sss)^2-\zz(\sss'^2-\sss^2)\bigr) + \Ord(\es^3).
\end{align}
Similarly, using the expansion of the exponential, we have
%Hence, using that $e^{\es (\sss-\sss')} = 1 + \es (\sss-\sss') + \frac{1}{2}\es^2 (\sss-\sss')^2 + \Ord(\es^3)$, we have
\begin{align}
    e^{\es (\sss-\sss')}\frac{\Bigl(1+\frac{\es\sss'}{\ew}\Bigr)^\zz}{\Bigl(1+\frac{\es\sss}{\ew}\Bigr)^\zz} % &= \left(1 + \es (\sss-\sss') + \frac{1}{2}\es^2 (\sss-\sss')^2 + \Ord(\es^3)\right)\left(1+\frac{\es}{\ew}\zz(\sss'-\sss)+\frac{\es^2}{2\ew^2}\bigl(\zz^2(\sss'-\sss)^2-\zz(\sss'^2-\sss^2)\bigr) + \Ord(\es^3)\right)\\
    &= 1 + \es\left(\frac{\zz}{\ew}-1\right)(\sss'-\sss) + \frac{\es^2}{2}\left(\left(1-\frac{\zz}{\ew}\right)^2(\sss-\sss')^2  + \frac{\zz(\sss^2-\sss'^2)}{\ew^2}\right) + \Ord(\es^3).
\end{align}
Now, carrying out the expectation over $x'$ we obtain
\begin{align}
    \sum_{\sss'\in\Xcal}\pmf(\sss')e^{\es (\sss-\sss')}\frac{\Bigl(1+\frac{\es\sss'}{\ew}\Bigr)^\zz}{\Bigl(1+\frac{\es\sss}{\ew}\Bigr)^\zz} &= 1 + \es\left(\frac{\zz}{\ew}-1\right)(\muOne-\sss) + \frac{\es^2}{2}\left(\left(1-\frac{\zz}{\ew}\right)^2(\sss^2+\muTwo-2\sss\muOne)  + \frac{\zz(\sss^2-\muTwo)}{\ew^2}\right) + \Ord(\es^3).
\end{align}

Next, using the expansion of the logarithm, %formula
%\begin{equation}
%    $\log(1+t) = t -\frac{1}{2}t^2+ \Ord(t^3)$,
%\end{equation}
we obtain
\begin{align}
    \log\left(\sum_{\sss'\in\Xcal}\pmf(\sss')e^{\es (\sss-\sss')}\frac{\Bigl(1+\frac{\es\sss'}{\ew}\Bigr)^\zz}{\Bigl(1+\frac{\es\sss}{\ew}\Bigr)^\zz}\right) %&=\es\left(\frac{\zz}{\ew}-1\right)(\muOne-\sss) + \frac{\es^2}{2}\left(\left(1-\frac{\zz}{\ew}\right)^2(\sss^2+\muTwo-2\sss\muOne)  + \frac{\zz(\sss^2-\muTwo)}{\ew^2}\right) -\frac{\es^2}{2}\left(\frac{\zz}{\ew}-1\right)^2(\muOne-\sss)^2+ \Ord(\es^3) \\
    &=\es\left(\frac{\zz}{\ew}-1\right)(\muOne-\sss) + \frac{\es^2}{2}\left(\left(1-\frac{\zz}{\ew}\right)^2(\muTwo-\muOne^2)  + \frac{\zz(\sss^2-\muTwo)}{\ew^2}\right) + \Ord(\es^3).
\end{align}

Now, multiplying by the channel law, we get for given $\sss$ and $\zz$
\begin{align}
%  \frac{\ew^\zz}{\zz!}&e^{-\ew}\left(\es\left(\frac{\zz}{\ew}-1\right)(\muOne-\sss) + \frac{\es^2}{2}\left(\left(1-\frac{\zz}{\ew}\right)^2(\muTwo-\muOne^2+2\muOne\sss-2\sss^2)  + \frac{\zz(\sss^2-\muTwo)}{\ew^2}\right)+ \Ord(\es^3)\right) \\
%  &= 
\frac{\ew^\zz}{\zz!}e^{-\ew}\left(\es\left(\frac{\zz}{\ew}-1\right)(\muOne-\sss) + \frac{\es^2}{2}\left(\left(1-\frac{\zz}{\ew}\right)^2(\muTwo-\muOne^2+2\muOne\sss-2\sss^2)  + \frac{\zz(\sss^2-\muTwo)}{\ew^2}\right)+ \Ord(\es^3)\right).
\end{align}
After carrying out the expectation over $\sss$, some terms cancel to give
\begin{align}
  \frac{\ew^\zz}{\zz!}e^{-\ew}\left(\frac{\es^2}{2}\left(1-\frac{\zz}{\ew}\right)^2(\muOne^2-\muTwo)+ \Ord(\es^3)\right).
\end{align}

As a final step, we sum over $\zz$ to obtain the mutual information,
\begin{align}
  \sum_{\zz=0}^\infty\frac{\ew^\zz}{\zz!}e^{-\ew}\left(\frac{\es^2}{2}\left(1-\frac{2\zz}{\ew}+\frac{\zz^2}{\ew^2}\right)(\muOne^2-\muTwo)+ \Ord(\es^3)\right) &= \frac{\es^2}{2\ew}(\muOne^2-\muTwo) + \Ord(\es^3).
\end{align}
%Therefore, the CM capacity is given by
%\begin{align}
%    \capCM &= \frac{\es^2}{2\ew}(\muTwo-\muOne^2) + \Ord(\es^3).
%\end{align}

\section{Asymptotics at Low $\es$ for Geometric Noise}
\label{app:cmExpansionAEQ}

The mutual information is given by
\begin{align}\label{eq:349}
    I(X;Y) &=  -\sum_\sss\pmf(\sss)\sum_{\zz=0}^\infty\pzx \log \Biggl(\frac{\sum_{\sss'\in\Xcal}\pmf(\sss')\pzxp}{\pzx}\Biggr),
\end{align}
where $\pzx$ is given by Eq.~\eqref{eq:chanzs-AEQ}.

As it happened in the discrete-time Poisson channel, the Taylor expansion of the exponential implies that there are only three possible channel outputs $\xx$ to order $\es^3$, that is,
\begin{align}
    &\xx = 0,\quad \pmf(\xx|\sss) = 1-\es \sss +\frac{1}{2}(\es \sss)^2+ \Ord(\es^3) \\
    &\xx = 1,\quad \pmf(\xx|\sss) = \es \sss -(\es \sss)^2 +\Ord(\es^3) \\
    &\xx = 2,\quad \pmf(\xx|\sss) = \frac{1}{2}(\es \sss)^2 + \Ord(\es^3) \\
    &\xx > 2,\quad \pmf(\xx|\sss) = \Ord(\es^3).
\end{align}
Hence the channel output $\zz = \xx + \ww$ only includes these contributions. We distinguish three cases, viz.\ $\zz = 0$, $\zz = 1$, and $\zz\geq 2$.

In the first case, $\zz = \xx = \ww = 0$, and $\pzx$ becomes
\begin{align}
    \pzx = \frac{1}{1+\ew}\Biggl(1-\es \sss +\frac{1}{2}(\es \sss)^2+\Ord(\es^3)\Biggr).
\end{align}

For $\zz = 1$, we combine the outputs $\xx = 0$ and $\xx = 1$,
%\begin{align}
%    \pzx &= \biggl(1-\es \sss +\frac{1}{2}(\es \sss)^2\biggr)\frac{1}{1+\ew}\biggl(\frac{\ew}{1+\ew}\biggr) + \frac{\es \sss -(\es \sss)^2}{1+\ew} +\Ord(\es^3).
%\end{align}
%Extracting a common term $\pmf(\ww = 1) = \ew/(1+\ew)^2$, which we will recover later in the analysis,
\begin{align}
    \pzx %&= \frac{\ew}{(1+\ew)^2}\Biggl(1-\es \sss +\frac{1}{2}(\es \sss)^2 + \frac{1+\ew}{\ew}\bigl(\es \sss -(\es \sss)^2\bigr) +\Ord(\es^3)\Biggr)\\
    &= \frac{\ew}{(1+\ew)^2}\Biggl(1+\frac{\es}{\ew} \sss -\es^2 \sss^2\biggl(\frac{1}{2} +\frac{1}{\ew}\biggr) +\Ord(\es^3)\Biggr).\label{eq:358}
\end{align}

For $\zz \geq 2$, we combine the outputs $\xx = 0$, $\xx = 1$, and $\xx = 2$,
%\begin{align}
%    \pzx &= \biggl(1-\es \sss +\frac{1}{2}(\es \sss)^2\biggr)\frac{1}{1+\ew}\biggl(\frac{\ew}{1+\ew}\biggr)^{\zz} + \frac{\es \sss -(\es \sss)^2}{1+\ew}\biggl(\frac{\ew}{1+\ew}\biggr)^{\zz-1} \notag \\
%    &\qquad + \frac{\frac{1}{2}(\es \sss)^2}{1+\ew}\biggl(\frac{\ew}{1+\ew}\biggr)^{\zz-2} +\Ord(\es^3).
%\end{align}
%As we just did for $\zz = 1$, we extract a common term $\ew^\zz/(1+\ew)^{\zz+1}$, which we will recover later in the analysis, and obtain
\begin{align}
    \pzx %&=\frac{\ew^\zz}{(1+\ew)^{\zz+1}}\Biggl(1-\es \sss +\frac{1}{2}(\es \sss)^2 + \bigl(\es \sss -(\es \sss)^2\bigr)\biggl(\frac{1+\ew}{\ew}\biggr) \notag \\
%    &\qquad\qquad\qquad\qquad + \frac{1}{2}(\es \sss)^2\biggl(\frac{1+\ew}{\ew}\biggr)^{2} +\Ord(\es^3)\Biggr) \\
%%    &= 1+\es \sss\biggl(-1+\frac{1+\ew}{\ew}\biggr) +\es^2 \sss^2\biggl(\frac{1}{2}-\frac{1+\ew}{\ew}+\frac{1}{2}\biggl(\frac{1+\ew}{\ew}\biggr)^{2}\biggr) +\notag \\
%%    &\qquad + \Ord(\es^3) \\
    &= \frac{\ew^\zz}{(1+\ew)^{\zz+1}}\Biggl(1+\frac{\es}{\ew} \sss +\frac{\es^2}{2\ew^2} \sss^2 + \Ord(\es^3)\Biggr),\label{eq:360}
\end{align}
after combining some terms together.

We next rewrite the numerator and denominator in the $\log(\cdot)$ in Eq.~\eqref{eq:349} with the appropriate approximation. For $\zz = 0$, the common term $(1+\ew)^{-1}$ cancels, and the numerator is
\begin{align}
    \sum_{\sss'\in\Xcal}{\pzxp\pmf(\sss')} %&= \sum_{\sss'\in\Xcal}\bigl(1-\es \sss' +\frac{1}{2}(\es \sss')^2 + \Ord(\es^3)\bigr)\pmf(\sss') \\
    &= \biggl(1 - \es\muOne+\frac{1}{2}\es^2 \muTwo\biggr) + \Ord(\es^3).\label{eq:362-2}
\end{align}
In the denominator, we keep the expansion
\begin{equation}
    1-\es \sss +\tfrac{1}{2}(\es \sss)^2\Ord(\es^3).\label{eq:363-2}
\end{equation}
Taking logarithms of Eqs.~\eqref{eq:362-2} and~\eqref{eq:363-2}, using a Taylor expansion, %formula
%%\begin{equation}
%    $\log (1+t) = t - \frac{1}{2}t^2 + \Ord(t^3)$,
%%\end{equation}
and combining numerator and denominator, we obtain
\begin{align}
    \log &\biggl(1 - \es\muOne+\frac{1}{2}\es^2 \muTwo+ \Ord(\es^3)\biggr) -\log\biggl(1-\es \sss +\frac{1}{2}(\es \sss)^2+\Ord(\es^3)\biggr) \\
%    \log &\bigl(1 - \es + \Ord(\es^2)\bigr) - \log \pmf(\sss) - \log \bigl(1 - \es \sss + \Ord(\es^2)\bigr) \\
%    &= - \es\muOne +\frac{1}{2}\es^2 \muTwo -\frac{1}{2}\es^2\muOne^2  +  \es \sss - \frac{1}{2}(\es \sss)^2+\frac{1}{2}(\es \sss)^2+ \Ord(\es^3) \\
    &= - \es(\muOne-\sss) +\frac{1}{2}\es^2 (\muTwo-\muOne^2)+ \Ord(\es^3) \label{eq:356}.
\end{align}

For $\zz = 1$, we use Eq.~\eqref{eq:358}. Summing over $\sss'$ in the numerator, we get
\begin{align}
    \sum_{\sss'\in\Xcal}{\pzxp\pmf(\sss')} &= \Biggl(1+ \frac{\es}{\ew}\muOne - \biggl(\frac{1}{2}+\frac{1}{\ew}\biggr)\muTwo\es^2 + \Ord(\es^3)\Biggr),
\end{align}
with the agreement that a common term $\ew/(1+\ew)^2$ has  been cancelled.

Combining numerator and denominator, taking logarithms, and using the Taylor expansion of the logarithm, we obtain
\begin{align}
     &\frac{\es}{\ew}\muOne - \biggl(\frac{1}{2}+\frac{1}{\ew}\biggr)\muTwo\es^2 - \frac{\es^2}{2\ew^2}\muOne^2- \frac{\es}{\ew} \sss + \biggl(\frac{1}{2}+\frac{1}{\ew}\biggr)\sss^2\es^2 +\frac{\es^2}{2\ew^2} \sss^2 + \Ord(\es^3) \label{eq:361} \\
%    &= \frac{(\muOne-\sss)\es}{\ew} - \Biggl(\biggl(\frac{1}{2}+\frac{1}{\ew}\biggr)\muTwo+ \frac{\muOne^2}{2\ew^2}-\biggl(\frac{1}{2}+\frac{1}{\ew}+\frac{1}{2\ew^2}\biggr)\sss^2\Biggr)\es^2 + \Ord(\e\sss^3) \label{eq:362} \\
    &= \frac{(\muOne-\sss)\es}{\ew} - \Biggl(\biggl(\frac{1}{2}+\frac{1}{\ew}\biggr)\muTwo+ \frac{\muOne^2}{2\ew^2}-\frac{\sss^2(1+\ew)^2}{2\ew^2}\Biggr)\es^2 + \Ord(\es^3) \label{eq:363}.
\end{align}

If the output is $\zz \geq 2$, in an analogous way we use Eq.~\eqref{eq:360} to rewrite the logarithm of the ratio of numerator and denominator as
\begin{align}
    \log\Biggl(1+\frac{\es}{\ew}\muOne  + \frac{\es^2}{2\ew^2} \muTwo  + \Ord(\es^3)\Biggr)-\log\Biggl(1+\frac{\es}{\ew} \sss +\frac{\es^2}{2\ew^2} \sss^2 + \Ord(\es^3)\Biggr).
\end{align}
Using now the Taylor expansion of the logarithm, we obtain
\begin{align}
%    &\frac{\es}{\ew}\muOne + \frac{\es^2}{2\ew^2} \muTwo - \frac{\es^2}{2\ew^2}\muOne^2 - \frac{\es}{\ew} \sss -\frac{\es^2}{2\ew^2} \sss^2 +\frac{\es^2}{2\ew^2}\sss^2 +  \Ord(\es^3) \label{eq:364} \\
%    &= 
\frac{\es}{\ew}(\muOne-\sss) + \frac{\es^2}{2\ew^2} (\muTwo  - \muOne^2) +  \Ord(\es^3) \label{eq:365}.
\end{align}

The remaining steps are the averaging over $\sss$ and $\zz$. We first carry out the expectation over $\sss$. From Eq.~\eqref{eq:356}, the averaging over $\sss$ yields
\begin{align}
  \sum_\sss\frac{1}{1+\ew}&\biggl(1-\es \sss +\frac{1}{2}(\es \sss)^2\biggr)\biggl((\sss-\muOne)\es +\frac{1}{2}\es^2 \bigl(\muTwo-\muOne^2\bigr) \biggr)+  \Ord(\es^3) \\
%  &= \frac{1}{1+\ew}\biggl(\frac{1}{2}\es^2 \bigl(\muTwo-\muOne^2\bigr)-\es^2\bigl(\muTwo-\muOne^2\bigr)\biggr)+  \Ord(\es^3)\\
  &= \frac{1}{1+\ew}\frac{1}{2}\es^2 \bigl(\muOne^2-\muTwo\bigr)+  \Ord(\es^3).\label{eq:5.164}
\end{align}

Similarly, from Eq.~\eqref{eq:363} we obtain (bar for a constant factor $\frac{\ew}{(1+\ew)^2}$)
\begin{align}
    \sum_\sss&\Biggl(1+\frac{\es}{\ew} \sss -\es^2 \sss^2\biggl(\frac{1}{2} +\frac{1}{\ew}\biggr)\Biggr)\times \notag \\
    &\qquad \times\Biggl(\frac{(\muOne-\sss)\es}{\ew} - \Biggl(\biggl(\frac{1}{2}+\frac{1}{\ew}\biggr)\muTwo+ \frac{\muOne^2}{2\ew^2}-\frac{\sss^2(1+\ew)^2}{2\ew^2}\Biggr)\es^2\Biggr) + \Ord(\es^3)\\
%   &= \Biggl(-\biggl(\frac{1}{2}+\frac{1}{\ew}\biggr)\muTwo- \frac{\muOne^2}{2\ew^2}+\frac{\muTwo(1+\ew)^2}{2\ew^2}+\frac{\es^2}{\ew^2}(\muOne^2-\muTwo)\Biggr)\es^2 + \Ord(\es^3) \\
   &= (\muOne^2-\muTwo)\frac{\es^2}{2\ew^2} + \Ord(\es^3).\label{eq:5.167}
\end{align}

And finally, from Eq.~\eqref{eq:365}, for $\zz\geq 2$, we get
\begin{align}
  \sum_\sss\frac{\ew^{\zz}}{(1+\ew)^{\zz+1}}&\biggl(1+\frac{\es}{\ew} \sss +\frac{\es^2}{2\ew^2} \sss^2\biggr)\biggl(\frac{\es}{\ew}(\muOne-\sss) + \frac{\es^2}{2\ew^2} (\muTwo-\muOne^2)\biggr)+\Ord(\es^3) \\
%  &=\frac{\ew^{\zz}}{(1+\ew)^{\zz+1}}\biggl(\frac{\es^2}{2\ew^2} (\muTwo  - \muOne^2) +  \frac{\es^2}{\ew^2} (\muOne^2-\muTwo)\biggr)+\Ord(\es^3) \\
  &=\frac{\ew^{\zz}}{(1+\ew)^{\zz+1}}\frac{\es^2}{2\ew^2} (\muOne^2-\muTwo)+\Ord(\es^3).\label{eq:5.170}
\end{align}

The summation over $\zz\geq 2$ can be carried out and yields
\begin{align}\label{eq:5.171}
    \sum_{\zz=2}^\infty \frac{\ew^{\zz}}{(1+\ew)^{\zz}} = \frac{\ew^2}{1+\ew}.
\end{align}
Then, combining Eq.~\eqref{eq:5.171} into Eq.~\eqref{eq:5.170}, and summing with Eqs.~\eqref{eq:5.164} and~\eqref{eq:5.167} (including the factor $\frac{\ew}{(1+\ew)^2}$), we obtain
\begin{align}
  \frac{1}{1+\ew}&\frac{1}{2}\es^2 \bigl(\muOne^2-\muTwo\bigr)\Biggl(1+\frac{1}{\ew(1+\ew)}+\frac{1}{\ew^2}\frac{\ew^2}{1+\ew}\Biggr)+  \Ord(\es^3) \\
  &=\frac{1}{2}\es^2 \bigl(\muOne^2-\muTwo\bigr)\frac{1}{\ew}+  \Ord(\es^3).
\end{align}
The expansion for $I(X;Y)$ follows.
%And, finally, the expansion for $\capCM$ follows,
%\begin{align}
%    \capCM &=  \frac{1}{2}(\muTwo-\muOne^2)\frac{\es^2}{\ew}+\Ord(\es^3).
%\end{align}


%\bibliographystyle{IEEEtran}
%\bibliography{IEEEabrv,amartine}

\end{document}